\documentclass[11pt,a4paper]{article}
\pdfoutput=1

\usepackage{jcappub}

\usepackage[latin1]{inputenc}
\usepackage{float}
\usepackage{amsmath,amssymb}
\usepackage{mathrsfs}
\usepackage{graphicx}
\usepackage{caption}
\usepackage{subcaption}
\usepackage{color}
\usepackage{wasysym}
\usepackage{hyperref} 
\usepackage{amsthm}

\definecolor{blue}{rgb}{0,0,1}
\definecolor{green}{rgb}{0,0.65,0.5}
\definecolor{red}{rgb}{1,0,0}
\definecolor{vio}{rgb}{0.66,0,1}
\definecolor{ama}{rgb}{1,1,0}
\definecolor{marron}{rgb}{0.7,0.2,0.1}

\title{ 
Geometrical models for the study of astrophysical systems with spheroidal symmetry
imbedded in a standard cosmology: 
The case of cosmic voids.
}

\author{Ezequiel F. Boero}
\author{and Osvaldo M. Moreschi}
\affiliation{
Facultad de Matem\'atica Astronom\'ia, F\'isica y Computaci\'on,
FaMAF, \\ Universidad Nacional de C\'{o}rdoba, \\
Instituto de F\'\i{}sica Enrique Gaviola, IFEG, CONICET,\\
Ciudad Universitaria, (5000) C\'{o}rdoba, Argentina.
 }

\emailAdd{boero@famaf.unc.edu.ar}
\emailAdd{moreschi@fis.uncor.edu}

\abstract{
We present a class of general prolate and oblate spheroidal spacetimes for the 
description of cosmic structures in the Universe. 
They are exact geometries which represent, in an appropriated way, the 
imbedding of spheroidal matter-energy distributions within a standard cosmological 
scenario, and therefore they allow for an 
improved description of a wider class of astrophysical systems from a 
more accurate point of view.
These spacetimes can be used to describe overdensity or underdensity regions;
in this work we consider the last case, that is, the description of cosmic 
voids. 
We introduce and study a model of void  which is a 
generalization of a simpler one in spherical symmetry and we use it for the calculation 
of weak lensing optical scalars as a non-trivial and interesting application.
In this particular example we show the rich observable features that can be
found in such models.

}


\arxivnumber{1611.05832}

\begin{document}
	\maketitle   
	\flushbottom 

\section{Introduction}

In the theoretical modelling of structures of the cosmic matter distribution,
the use of spherically symmetric models has been the first natural option employed
for the study of such systems.
Notably, several astrophysical objects such as galaxy clusters and most cosmic voids 
are described using this option (See for example \cite{Hamaus:2014fma, Inoue:2006fn}). 
In an standard cosmology, there exists a reason for considering spherically symmetric 
models; namely, that a priori, in a Robertson-Walker (R-W) background it is natural 
to expect to find structures with spherical shape since no preferred directions exist 
in such geometry. 
However, in our lumpy Universe a great variety of shapes are observed; this
fact mainly concerns to the very relevant case of voids.

Cosmic voids are defined as underdensity regions with respect to the mean cosmic
energy density $\bar{\varrho}$ of the Universe and they comprise one of the most 
interesting large scale objects due to its predominant presence in the Universe.
Analysis of the 2dF Galaxy Redshift Survey and the Sloan Sky Digital Survey
\cite{Hoyle:2003hc, Pan:2011hx} have shown that most part of the Universe is filled out 
with voids; and about the 40\% of its total volume is occupied by voids with radius of 
the order of 15Mpc \cite{Hoyle:2003hc}.

Although very often they do not present very well defined shapes,
it have been suggested that a more appropriated characterization of the 
morphology of voids can be obtained by means of ellipsoids 
\cite{Foster:2009rt, Tikhonov:2006bv, vandeWeygaert:2009hr}; though
in most works the algorithms that search for voids had assumed spherical symmetry. 	 

Dealing with models containing deviations from spherical symmetry confronts
the researcher with a considerable increasing of complexity in the 
description of the system; 
nevertheless there exists analytic studies\cite{Park:2006wu} using these geometries.
However, works found in the literature
avoid the use of a detailed metric description for the spacetime of such
ellipsoidal 
configurations inferred from observations. 
Such kind of models are necessary for several reasons; for example it would allow 
a more realistic characterization of voids and at the same time they offer the possibility
of explore more general energy contents in terms of its whole energy-momentum tensor.
This is very interesting taking into account that the nature of the dark sector
is actually unknown and that most of the models rest on Newtonian descriptions.
Additionally, having exact metrics that describe cosmic voids is relevant
for the study of the dynamical aspects of underdensity regions
in the problem of evolution and formation of structure in the Universe.

Unfortunately, there is a lack of suitable metrics at our disposal for describing 
ellipsoidal configurations and therefore the amount of work regarding most sophisticated 
modeling and its implications is not abundant within the literature.

This work is intended to be a contribution in these directions.
Here we present a family of exact metrics in order to model 
spacetimes of large structures with prolate and oblate
spheroidal symmetry imbedded in a standard Robertson-Walker (R-W) cosmology. 
This means that they are non-perturbative models, and furthermore, 
they possess two well distinguished regions, 
namely an inner region which describes a very 
general class of spheroidal distributions and an outer regions 
which is chosen to coincide 
appropriately with an exact isotropic and homogeneous metric. 
Therefore, they constitute a straightforward generalization of spherically 
symmetric models surrounded by a cosmological environment.
The metrics presented in this article can be used to describe
large scale structure as galaxy clusters and cosmic voids; 
however we will deal in this case with underdensity distributions.

In particular, we analyse below voids with a mass density 
profile which is an adapted version of a simple model\cite{Amendola1999} 
used in spherical symmetry for the study of the probability of detection 
of weak lensing in the presence of voids.
Inspired by reference \cite{Amendola1999}, we use this particular geometric
model of a void in a weak 
lens calculation of the optical scalars in the regime of thin lenses in which 
we also present an 
application of the recently reported new expressions for 
the lens optical scalars in the cosmological context \cite{Boero:2016nrd}.

	We have not found in the literature works that use exact spheroidal metrics
in a cosmological context to describe voids.
Although these geometries constitute a slight modification 
to spherically symmetric systems, they become quite sophisticated since, for example,
the model
also allows for non-vanishing spacelike components of the energy-momentum tensor.
The reason behind considering this broader assumptions in the degrees 
of freedom of the matter
content comes from the fact that
there exist interesting suggestions\cite{Gallo:2011hi} that such flexibility in the 
energy content is a valuable alternative to give new insight in the study of the 
matter content of cosmological and astrophysical systems. 
This, certainly concerns to voids which are supposed to be one
of the typical candidates for the study of the distribution 
of the cosmic mass-energy content; 
see for example \cite{Gottloeber:2003zb, Melchior:2013gxd}. 

Almost all of the information that we have gather for building our
cosmological picture, comes from the present past null cone.
That is, from each system we are seeing a snapshot.
One of the objectives of this work is to communicate a very convenient
way to build geometrical models for observed astrophysical system;
since it can be shown that the choice of two functions can be
used to model a variety of systems with spheroidal symmetry
immersed in a cosmological scenario.
We will not concentrate in the detailed physical sources
of these geometries, but the functions can be chosen so that
the usual energy conditions are satisfied\cite{Gallo:2011hi}.

This article is organized as follows.
In the next section we recall some elementary geometric relations that 
will be used in the rest of the article.
Section \ref{sec:exact} presents the line element of the family of models 
with different spheroidal symmetry imbedded in a Robertson-Walker cosmology.
In section \ref{sec:voidmodel} we apply this construction to the modeling 
of a cosmic void
by presenting a metric that have the appropriate geometric behaviour.
In section \ref{sec:weak} we present the results of the calculation of
the optical scalars of this spheroidal void model; which is considered
inclined along the line of sight.

\section{Geometric Preliminaries}

\subsection{Spheroids in Euclidean space}
There are two natural classes of spheroidal coordinates systems, namely prolate and oblate,
and they are generated from elliptic coordinates in the Cartesian plane $\mathbb{R}^2$ 
\cite{moon1988field}.
Spheroidal coordinates are normally chosen so that its radial coordinate describe a set 
of confocal spheroids.
Prolate spheroidal coordinates are obtained by rotating the confocal ellipses level set 
along the axis that join the two foci; while that oblate coordinates systems are generated 
rotating the ellipses level set along an axis perpendicular to the confocal one; this means 
that after the rotation, the foci draw a ring in the space.

In order to fix the notation,
below we introduce both coordinates systems in the usual way trough its relation with  
Cartesian coordinates $\left(\mathbf{x}, \mathbf{y}, \mathbf{z} \right)$ in Euclidean space 
$\mathbb{R}^3$.

\subsubsection{Prolate spheroidal coordinates}
We will denote by $\{ r, \theta, \phi\}$ the prolate spheroidal coordinates which have the 
following domain of definition:
\begin{align}
  r &\to [0, \infty),  \label{interval:r-range}\\
  \theta &\to [0, \pi ], \label{interval:theta-range} \\
  \phi &\to [0, 2\pi ]; \label{interval:phi-range}
\end{align}
and are related to the usual Cartesian coordinates $\left(\mathbf{x}, \mathbf{y}, \mathbf{z} \right)$
by:
\begin{align}
\mathbf{x} &= r \sin(\theta) \cos(\phi), \\
\mathbf{y} &= r \sin(\theta) \sin(\phi), \\
\mathbf{z} &= \sqrt{r^2 + r^2_\mu} \cos(\theta);
\end{align}
where the constant parameter $r_\mu$ has the meaning of the distance of the 
foci from the origin of the Cartesian coordinate system.

One can see also that surfaces $r=$constant are prolate spheroids with their foci along the 
$\mathbf{z}-$axis, and therefore satisfy:
\begin{equation}
\frac{\mathbf{z}^2}{r^2 + r^2_{\mu}} + \frac{\mathbf{x}^2 + \mathbf{y}^2}{r^2}
= 1 ;
\end{equation}
where the value $r=constant$ correspond to the length of its minor radius and
the size of its major radius is equal to $\sqrt{r^2 + r^2_{\mu}}$. 

In this coordinates system the line element of the flat Euclidean metric in $\mathbb{R}^3$
acquires the following expression:
\begin{equation}\label{eq:ds-flat-3d-prolate}
ds^2 = \left( r^2 + r_{\mu}^2 \sin(\theta)^2 \right)\left( \frac{dr^2}{r^2 + r_{\mu}^2} 
+ d\theta^2 \right) + r^2 \sin(\theta)^2 \; d\phi^2.
\end{equation}

\subsubsection{Oblate spheroidal coordinates}

Oblate coordinates are usually labeled by the same set of symbols $\{ r, \theta, \phi\}$ 
employed in the prolate case, and additionally the range of this coordinates is also 
given by (\ref{interval:r-range}), (\ref{interval:theta-range}), (\ref{interval:phi-range}).

These coordinates are defined in terms of Cartesian coordinates by
\begin{align}
\mathbf{x} &= \sqrt{r^2 + r^2_\mu}  \sin(\theta) \cos(\phi), \\
\mathbf{y} &= \sqrt{r^2 + r^2_\mu} \sin(\theta) \sin(\phi), \\
\mathbf{z} &= r \cos(\theta).
\end{align}

The constant parameter $r_\mu$ in this case corresponds to the radius described by the
foci of the ellipse generating the spheroid when it is rotated around the $\mathbf{z}$-axis.

Let us note that the surfaces $r=$constant describe oblate spheroids with the $\mathbf{z}$ symmetry axis;
since one has:
\begin{equation}
\frac{\mathbf{z}^2}{r^2} + \frac{\mathbf{x}^2 + \mathbf{y}^2}{r^2 + r^2_{\mu}}
= 1;
\end{equation}
and where $\sqrt{r^2 + r^2_{\mu}}$ corresponds to the major radius and $r$ the minor one in 
completely analogy with the prolate case. 

The expression of the line element of the flat Euclidean space $\mathbb{R}^3$ in terms of
oblate coordinates becomes:
\begin{equation}\label{eq:ds-flat-3d-oblate}
\begin{split}
ds^2 = \big( r^2 + r_{\mu}^2 \cos(\theta)^2 \big)\left( \frac{dr^2}{r^2 + r_{\mu}^2} 
+ d\theta^2 \right) + \big( r^2 + r_{\mu}^2 \big)\sin(\theta)^2 d\phi^2.
\end{split}
\end{equation}

\subsection{Homogeneous and isotropic spacetime in spheroidal coordinates systems}

\subsubsection{The standard R-W line element}

Since the surrounding geometry of the spheroidal systems are required to be a 
standard R-W cosmology, before presenting our proposed geometric models 
we introduce here a description of the R-W metrics in terms of the
spheroidal coordinates discussed previously.
This form of the line element constitutes the base of the generalized models 
that we present below.

The line element of a Robertson-Walker Universe can be expressed by
\begin{equation}\label{eq:RW-metric}
ds^2 = dt^2 - A(t)^2 dL_k^2;
\end{equation}
where $A(t)$ is the expansion parameter and $dL^2_k$ denotes the line element of an 
homogeneous and isotropic space with constant curvature; where as usually 
$k=1, 0, -1$ refers to a positive curved, flat and hyperbolic spatial geometries 
respectively.
Among the several equivalent expressions for
the line element $dL^2_k$, we choose the following one which 
exhibits explicitly the fact that $dL^2_k$ is conformal to the Euclidean 3-dimensional 
flat metric
\begin{equation}\label{eq:dL-k-appropriated}
dL^2_k = \frac{1}{\left( 1 + \frac{k \bar{r}^2}{4}\right)^2}\left[ d\bar{r}^2 + 
\bar{r}^2 \Big( d\bar{\theta}^2 + \sin(\bar{\theta})^2 d\bar{\phi}^2 \Big)\right].
\end{equation}
Other usual ways in which $dL^2_k$ is often expressed are:
\begin{align}
dL^2_k =&  \frac{d\mathtt{r}^2}{1 - k \mathtt{r}^2} + 
\mathtt{r}^2 \Big( d\bar{\theta}^2 + \sin(\bar{\theta})^2 d\bar{\phi}^2 \Big), \\
dL^2_k =&  d\chi^2 + 
f_k(\chi)^2 \Big( d\bar{\theta}^2 + \sin(\bar{\theta})^2 d\bar{\phi}^2 \Big);
\end{align}
where in the last expression $f_k(\chi)$ is equal to $\sin(\chi)$ for $k=1$,
$\chi$ for $k=0$ or $\sinh(\chi)$ for $k=-1$.
However, we prefer to use equation (\ref{eq:dL-k-appropriated}) since we will exploit its
conformal nature using prolate and oblate coordinate system for writing
the flat metric inside of the brackets in order to describe spheroidal systems.

\subsubsection{The R-W line element in prolate spheroidal coordinates}
Using the line element (\ref{eq:ds-flat-3d-prolate}) together with the 
relation 
\begin{equation}
\bar{r}^2 \equiv \mathbf{x}^2 + \mathbf{y}^2 + 
\mathbf{z}^2 = r^2 + r^2_\mu \cos(\theta)^2;
\end{equation}
which holds for prolate coordinates we obtain for the R-W line element (\ref{eq:RW-metric}) 
\begin{equation}
\begin{split}\label{eq:prolateRW}
ds^2 = dt^2 - \frac{A(t)^2}{\left[ 1 + \frac{k}{4} \Big(r^2 + r^2_\mu \cos(\theta)^2 \Big) \right]^2}
\bigg[\left( r^2 + r_{\mu}^2 \sin(\theta)^2 \right) 
\left( \frac{dr^2}{r^2 + r_{\mu}^2} 
+ d\theta^2 \right) + r^2 \sin(\theta)^2 \; d\phi^2 \bigg].
\end{split}
\end{equation}

\subsubsection{The R-W line element in oblate spheroidal coordinates}
In this case it is immediate to see that the following relation holds 
\begin{equation}
\bar{r}^2 \equiv \mathbf{x}^2 + \mathbf{y}^2 + \mathbf{z}^2 = r^2 + r^2_\mu \sin(\theta)^2;
\end{equation}
which together with the line element (\ref{eq:ds-flat-3d-oblate}) implies
\begin{equation}
\begin{split}\label{eq:oblateRW}
ds^2 = dt^2 - \frac{A(t)^2}{\left[ 1 + \frac{k}{4} \Big(r^2 + r^2_\mu \sin(\theta)^2 \Big) \right]^2}
\bigg[\left( r^2 + r_{\mu}^2 \cos(\theta)^2 \right) 
\left( \frac{dr^2}{r^2 + r_{\mu}^2} 
+ d\theta^2 \right) + \big( r^2 + r^2_\mu \big) \sin(\theta)^2 \; d\phi^2 \bigg].
\end{split}
\end{equation}

It is probably worthwhile to emphasize that the line elements (\ref{eq:prolateRW}) and
(\ref{eq:oblateRW}) represent the same Robertson-Walker metric (homogeneous and isotropic) in
the less conventional spheroidal coordinate systems.

\section{Exact spacetime models for spheroidal systems in cosmology}\label{sec:exact}

Now, we are in condition to present our models for prolate and oblate spheroidal astrophysical 
systems imbedded in an standard R-W cosmology.


\subsection{Geometry with prolate symmetry}

It is possible to generalize the standard homogeneous and isotropic R-W spacetime
to one describing a distribution of matter with spheroidal symmetry.
This symmetry must be reflected in the geometry of the corresponding spacetime.
We do this by introducing two new functions that alter the R-W line element,
in a bounded region,
but that conforms to the spheroidal symmetry.

For the case of the prolate spheroidal symmetry the proposed line element is:


\begin{equation}\label{eq:ds-general-prolate-imbedded}
\begin{split}
ds^2 = F(r) dt^2 - 
\frac{A(t)^2}
{\left[ 1 + \frac{k}{4} \Big(r^2 + r^2_\mu \cos(\theta)^2 \Big) \right]^2} 
\Bigg[& \left( r^2 + r^2_\mu \sin(\theta)^2 \right) 
\Bigg( \frac{dr^2}{r^2 - \frac{2M(r)}{r} \Big(r^2 + r^2_\mu \sin(\theta)^2 \Big) + r^2_\mu} 
+ d\theta^2 \Bigg)  \\
&  + 
r^2 \sin(\theta)^2 d\phi^2  \Bigg] .
\end{split}
\end{equation}

\subsection{Geometry with oblate symmetry}

Applying the same methodology to the case of oblate spheroidal symmetry, we propose
the line element:

\begin{equation}\label{eq:ds-general-oblate-imbedded}
\begin{split}
ds^2 = F(r) dt^2 - \frac{A(t)^2}
{\left[ 1 + \frac{k}{4} \Big(r^2 + r^2_\mu \sin(\theta)^2 \Big) \right]^2} 
\Bigg[& \left( r^2 + r^2_\mu \cos(\theta)^2 \right)
\Bigg( \frac{dr^2}{r^2 - \frac{2M(r)}{r} \Big(r^2 + r^2_\mu \cos(\theta)^2 \Big) + r^2_\mu} 
+ d\theta^2 \Bigg)  \\ 
& + 
\left( r^2 + r^2_\mu \right) \sin(\theta)^2 d\phi^2  \Bigg].
\end{split}
\end{equation}

One observes the presence of the radial function $M(r)$ in the component $g_{rr}$ of 
the metric, which we call the \emph{mass profile function} due to its 
resemblance with the 
usual mass function in spherical symmetry, and the presence of 
the radial function, $F(r)$ 
which corresponds to the timelike component $g_{tt}$ of the metric.
One can see that the spacetime is still foliated by spacelike 
sections ($t=\text{constant}$) with spheroidal symmetry.
It is also clear that if we set $M(r)=0$ and $F(r)=1$ in a region specified 
by $r >r_0$ with 
$r_0=$constant, then, in this outer region,
the solution coincides exactly with a R-W geometry,
and therefore the dynamics of the scale factor $A(t)$ is determined 
by the field equations in this region.
In the interior region one has a wide range of possible models with spheroidal 
symmetry which can be adjusted in order to describe several matter-energy
distributions.

The mass profile function $M(r)$ is a  natural generalization of the 
mass function in spherical symmetry; in particular it gives the mass 
content in the limit when $r_\mu \to 0$. 

The function $F(r)$ is another degree of freedom that one can use to model the
physical system. It can be used to introduce non-negligible peculiar 
spacelike components to the energy-momentum that have shown to be useful for the description of
dark matter phenomena \cite{Gallo:2011hi}.

Therefore, the line elements (\ref{eq:ds-general-prolate-imbedded}) and
(\ref{eq:ds-general-oblate-imbedded}) constitute a non-trivial class of geometries 
for the practical study of spheroidal distributions in a cosmological context. 
We will show their utility in the special case of cosmic under-density regions; 
in particular 
we will regard voids from the point of view of weak gravitational lenses 
and we will perform 
the computation of the corresponding optical scalars.

\section{An exact geometrical model for spheroidal cosmological voids}\label{sec:voidmodel}
In this section we present a simple and concrete model for spheroidal voids; 
that is, a geometry whose associated energy-momentum tensor (via Einstein equations)
 has a 
timelike component, namely $T_{t}^{\;\,t}$, which in a given region $r < r_0$, 
is lower than the mean cosmic energy-density in the exterior region $r \geqslant r_0$.

If we consider the simple choice $F(r)=1$, then the candidate mass profile 
function $M(r)$ 
that one can consider is the one taken from the work \cite{Amendola1999} which 
is specified by:
\begin{equation}\label{eq:M-profile}
M(r) = 
\begin{cases}
\frac{4\pi}{3} \varrho_{\text{int}} r^3  & \text{for} \quad r < R \\
M(R) + \frac{4\pi}{3} \varrho_{\text{bor}} \left( r^3 - R^3 \right) 
& \text{for} \,  R \leq  r < R + d  \\
0 & \text{for} \quad R + d \leq  r ;
\end{cases}
\end{equation}
where $\varrho_{\text{int}}$ and $\varrho_{bor}$ are related to the value of mean 
density outside of the void, namely $\rho_0$ and the size parameters $R$ and $d$ in the 
following way:
\begin{align}
\varrho_{\text{int}} &= - \rho_0 \epsilon \qquad \qquad \qquad (\text{with} \quad \epsilon \leq 1), \\
\varrho_{\text{bor}} &= \frac{\rho_0 \epsilon}{\left( 1 + \frac{d}{R} \right)^3 - 1} 
\; \qquad (\text{with} \quad \epsilon \leq 1).
\end{align}

The parameter $R$ is associated to the `radius' of the void while the parameter $d$ 
 is related to the size of the wall, that is needed for the compensation.
In spherically symmetric cases, this profile is such that the mass at the border
compensates for the amount missing in the void and in this way one obtains a compensated void.
The value of the density $\rho_0$ outside of the voids corresponds to the mean cosmic 
density which according to the Friedman equation satisfies: 
\begin{equation}
\frac{3}{A(t)^2} \left[\left( \frac{dA}{dt}\right)^2 + 
k \right] = 8\pi \rho_0.
\end{equation}

\begin{figure}[H]
\centering
\includegraphics[clip, width=0.6\textwidth]{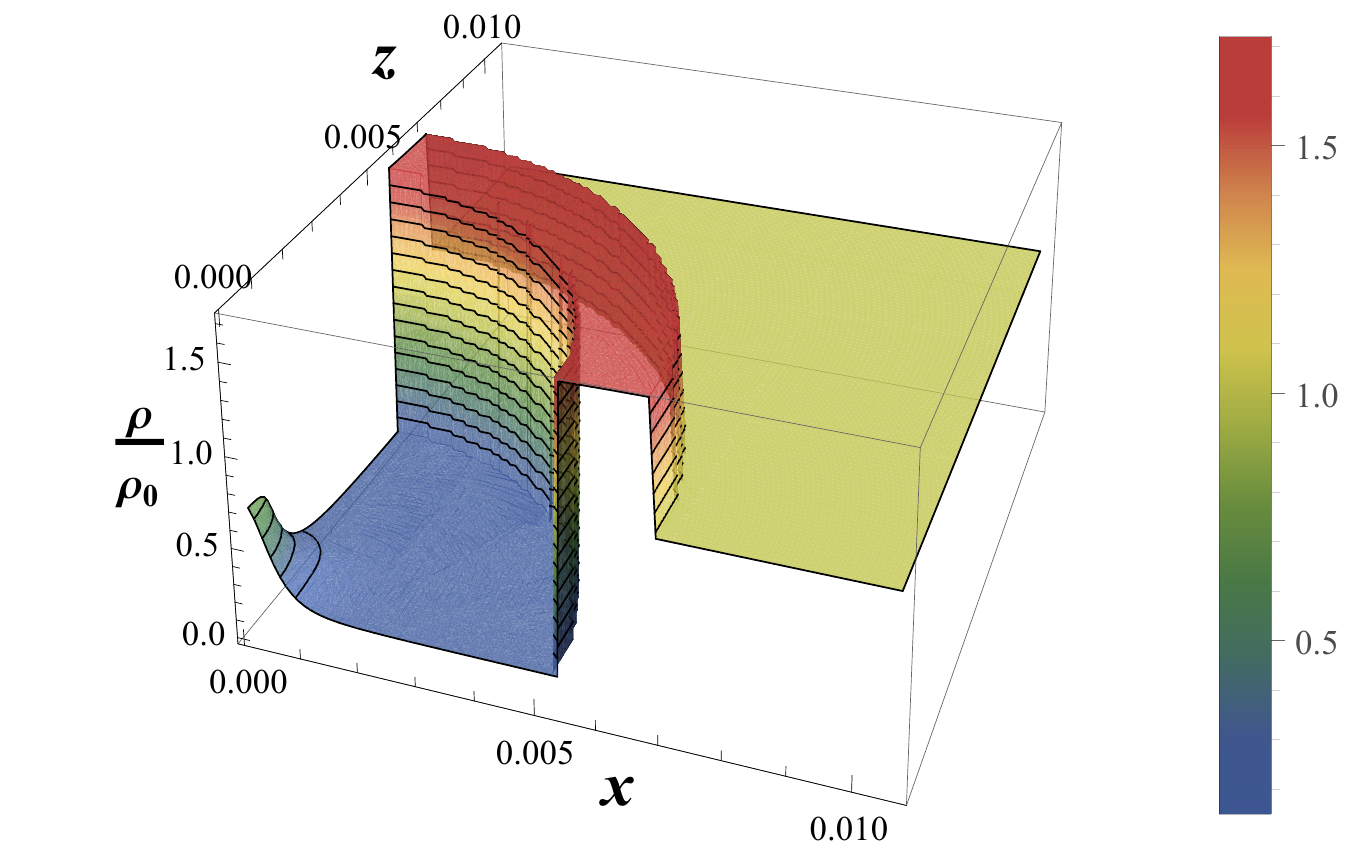}
\caption{
	Graphs of the relative energy-density with respect to the mean cosmic density, for a 
	compensated void with prolate spheroidal symmetry, imbedded in a Robertson-Walker spacetime;
	with symmetry axis along the $z$ direction. 
We have used the geometric parameters $d = 0.3 R$, $\epsilon = 0.9$,
$r_\mu = 0.1 R$, with a radius $R$ corresponding to a physical size of 22Mpc.
}
\label{fig:ProlateVoid}
\end{figure}

The explicit form of the components of the energy-momentum tensor are quite
involved; so that instead of presenting the complicated analytical form
of the energy momentum tensor, we show the graph of the energy density
in figure \ref{fig:ProlateVoid}.
It can be seen that our simplified model provides with a
a very good description of a region that can be
interpreted as an under-density compensated region.

\section{Weak gravitational lensing of spheroidal voids}\label{sec:weak}

\subsection{General thin lens equations}
In this section we present the numeric computation 
of the optical scalars when we regard the void as a weak thin lens.

Our calculations make use of the weak lensing formalism developed in 
\cite{Gallo11, Boero:2016nrd} where new expressions for the optical 
scalars describing 
the distortion of images by the weak gravitational field of a general distribution were
reported.
These expressions 
have the advantage of including the whole energy-momentum tensor in the 
description of the
lens in contrast to the formalism usually employed which is based in the 
linear superposition of 
massive point like deflectors.

In the cosmological context the convergence $\kappa$ and shear components 
$\left( \gamma_1 , \gamma_2 \right)$ of a thin lens are given by\cite{Boero:2016nrd}
\begin{align}
\kappa &= \left( 1 - \kappa_c  \right) \kappa_L
 + \kappa_c 
, \\
\gamma_{1} + i \gamma_{2} &= \left( 1 - \kappa_c  \right) 
(\gamma_{1L} + i \gamma_{2L})
;
\end{align}
with
\begin{align}
\kappa_L &= \textbf{D}_{ls} \int_{\lambda_l}^{\lambda_s} \delta\Phi_{00} \, d\lambda ,\\
\gamma_{1L} + i \gamma_{2L} &= \textbf{D}_{ls} \int_{\lambda_l}^{\lambda_s} \delta\Psi_{0} \,  d\lambda  ;
\end{align}
where $\delta \Phi_{00}$ and $ \delta\Psi_{0}$ represent the departure of the components
$\Phi_{00}$ and $\Psi_0$ of the Ricci tensor and the Weyl tensor 
from the R-W geometry;
$\lambda$ denotes the affine parameter along the null geodesics 
and the \emph{cosmological convergence} $\kappa_c$ together with 
the factor $\textbf{D}_{ls}$ are given by:
\begin{equation}\label{eq:kappa-c}
\kappa_c = 1 - \frac{D_A(\lambda)}{\lambda},
\end{equation}
\begin{equation}\label{eq:DLS}
\textbf{D}_{ls} = \frac{1}{1 + z_l} \frac{D_{A_{sl}} D_{A_l}}{D_{A_s}};
\end{equation} 
where $D_A$ denotes angular diameter distances and $z$ the cosmological redshift;
the subindex $s$ and $l$ refer to the source and the lens respectively
(For details we refer the reader to \cite{Boero:2016nrd}).

\subsection{Optical scalars for a prolate spheroidal void tilted with respect to the line of sight} 

\begin{figure}[H]
	\centering
	\begin{subfigure}[b]{0.25\textwidth}
		\centering
		\includegraphics[width=\textwidth]{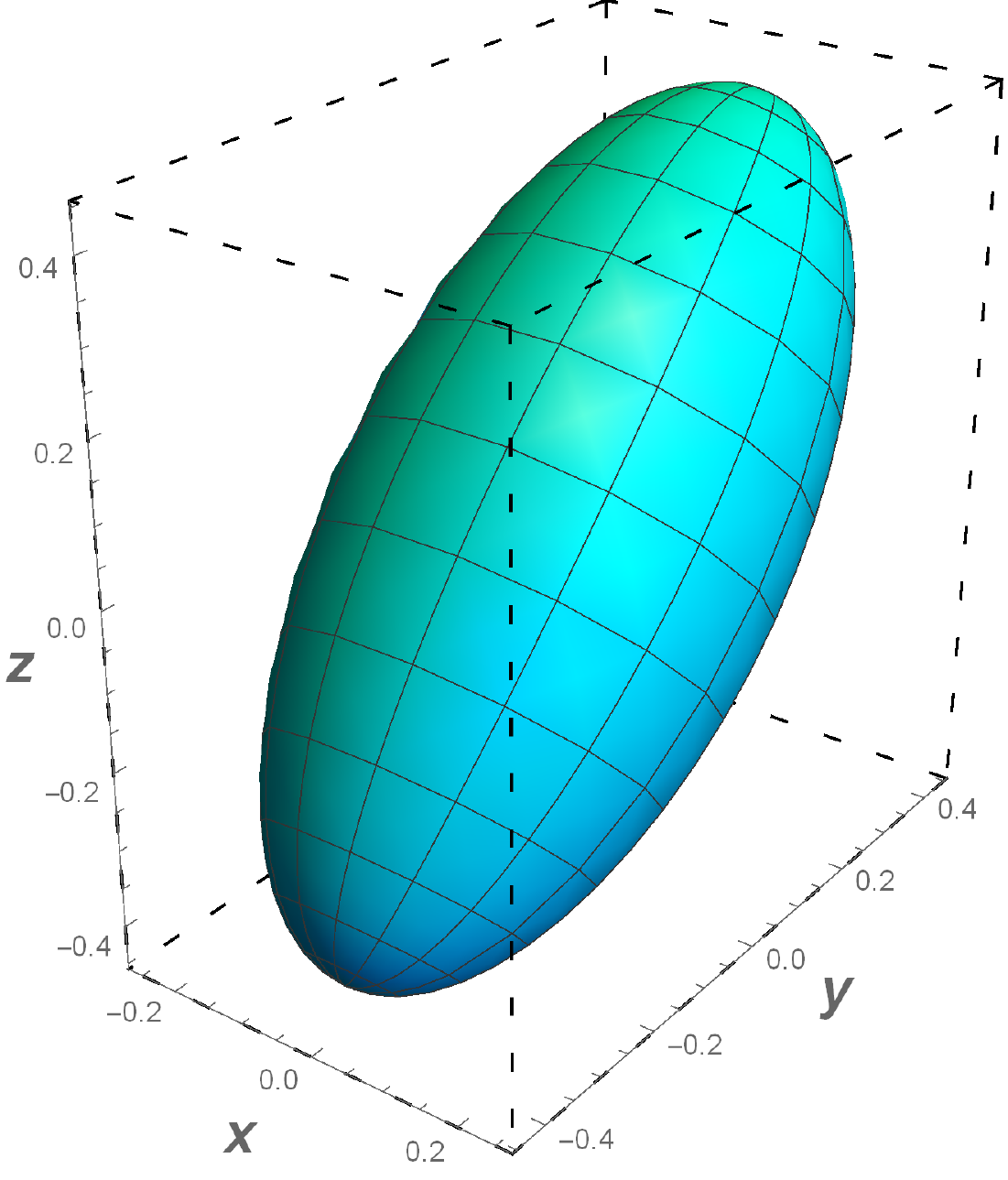}
		\label{fig:KappaL-Cluster}
	\end{subfigure}%
	\qquad \qquad \qquad 
	\begin{subfigure}[b]{0.25\textwidth}
		\centering
		\includegraphics[width=\textwidth]{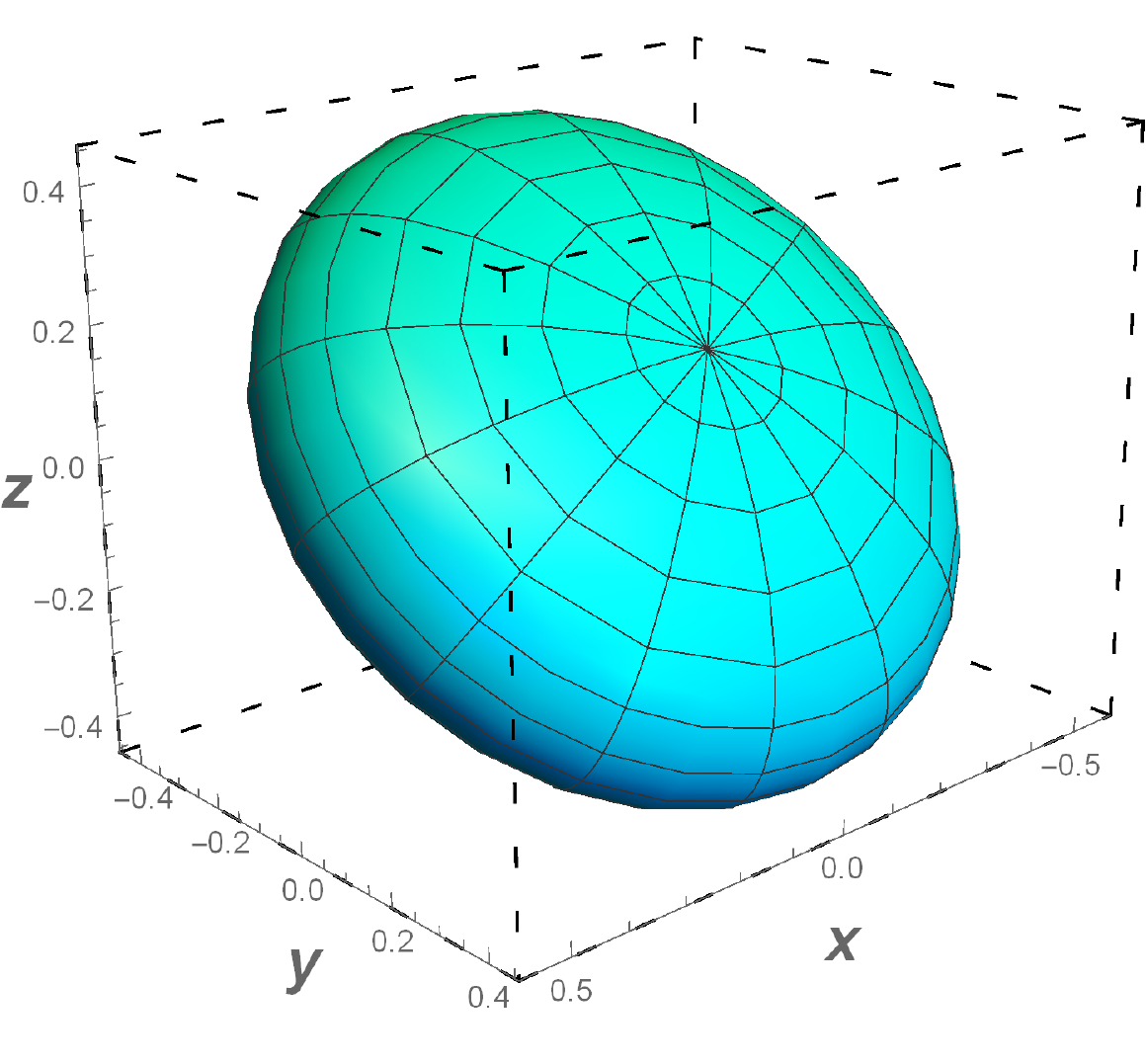}
		\label{fig:GammaL-Cluster}
	\end{subfigure}
~ 
\caption{Tilted prolate and oblate spheroids.
Photons travel approximately along the $y$ direction which is the line of sight of the 
observer.}
\label{fig:ProlateSpheroid}
\end{figure}

In our calculation we take into account the possibility that the spheroidal geometry is 
tilted with respect to the line of sight by an angle $\iota$. 
We adopt the convention that the plane of the lens which is locate at the
 cosmological distance $\lambda_l$, 
correspond to the plane $y=0$ in the local reference system associated to the spheroid in which 
it is 
inclined with respect to the vertical $z$-axis as it is show in the 
figure \ref{fig:ProlateSpheroid}.

Working in the thin lens approximation 
is consistent with the assumption that 
the time of flight of the photons, in the void, 
is negligible in comparison with the cosmological 
distance traveled in the R-W geometry;
therefore, for all practical purposes one can consider the geometry of the 
spheroidal lens as static; in other words, we assume that while the 
photons travel along the void,
the expansion parameter has the constant value $A_\mathbf{v}$.

In this section we will consider a timelike component of the metric 
$F(r) \neq 1$.
The function $F(r)$ that we have chosen is an adapted version of the one 
appearing in the timelike component of a spherically symmetric
peculiar geometry presented in the work \cite{Gallo:2011hi}.
Such functional form plays a key role in the description of many of the characteristic phenomenology 
observed in astrophysical system with dark matter\cite{Gallo:2011hi};
and it is given by 
\begin{equation}\label{eq:F-de-r-peculiar}
F(r) = F_0 \left[ C + \ln \left( \frac{r}{r_\mu} + \sqrt{1 + \frac{r^2}{r_\mu^2}}\right) \right]^2;
\end{equation}
where $F_0$ and $C$ are both constant parameters.
The main feature that introduces this particular functional dependence is 
that the spacelike
components of the energy-momentum tensor become non-negligible with respect to the 
timelike one and therefore it produces a significant contribution which is reflected 
in weak lensing calculations.

The following results correspond to a prolate spheroidal void which is tilted an angle
$\iota = \frac{\pi}{4}$ with respect to the line of sight.
It has a typical size such as those found in catalogs (as for example in \cite{Hoyle:2003hc}).
Below we give the detailed values used in this numerical calculation.
The parameter $R$ in equation (\ref{eq:M-profile}) was chosen so that the void has a typical 
radius of about 20Mpc, the parameter $d$ controlling the width of the walls has
been taken as $d=0.3R$, while $r_\mu$ corresponds to 3Mpc.
The parameter $\epsilon$, also involved in equation (\ref{eq:M-profile}) was taken equal 
to 0.9. 

The selection of the constants $C$ in (\ref{eq:F-de-r-peculiar}) was made  
ensuring that the component $T_{r}^{\;r}$ of the energy-momentum tensor 
remains of the same order as the mean density 
cosmic density $\rho_0$; in our 
case $C=10^7$.

The cosmology employed was a $\Lambda$CDM with cosmological parameters taken from the 
values reported by the Planck Collaboration \cite{Ade:2015xua} and the 
factor $\textbf{D}_{ls}$ was taken $\sim 220 \text{Mpc}$.

Below we show the graphs of the optical scalar $\kappa_L$, the components $\gamma_{1L}$ and
$\gamma_{2L}$ of the shear, and of its module $\gamma$ for a prolate spheroidal distribution.
The relation among the components of the shear and its module is expressed by
\begin{equation}\label{eq:shear-complejo-fase}
\gamma_1 + i \gamma_2 \equiv - \gamma e^{i2 \vartheta};
\end{equation}
where the phase $\vartheta$ has the meaning of an angle in the plane $x-z$ of the lens. 

In the case of spheroids the phase $\vartheta$ has a complicated relation in terms of the
impact position in the lens plane which can be seen by inspection of the individual 
components $\left( \gamma_1, \gamma_2 \right)$.


\begin{figure}[H]
		\centering
		\includegraphics[width=0.49\textwidth]{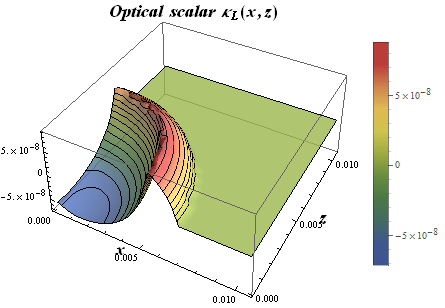}
		\includegraphics[width=0.49\textwidth]{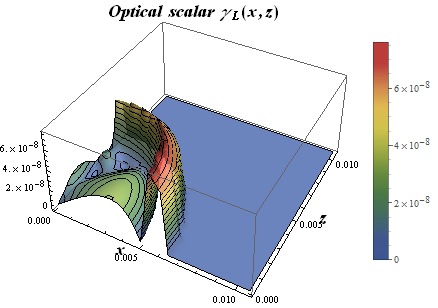}
	\caption{
		Graphs of the optical scalars in terms of coordinates of 
			the plane of the lens for a prolate spheroidal void tilted 
			$\iota = \frac{\pi}{4}$ respect to the line of sight.
		Left: the optical scalar parameter $\kappa$. 
		Right: the module of the shear $\gamma$. 
		}
	\label{fig:kappa-shear}
\end{figure}

\begin{figure}[H]
		\centering
		\includegraphics[width=0.49\textwidth]{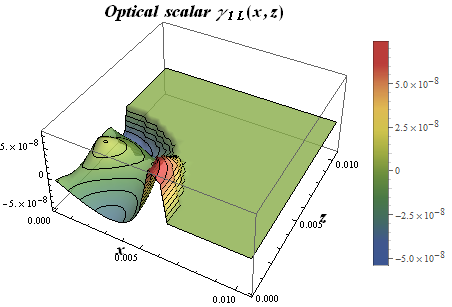}
		\includegraphics[width=0.49\textwidth]{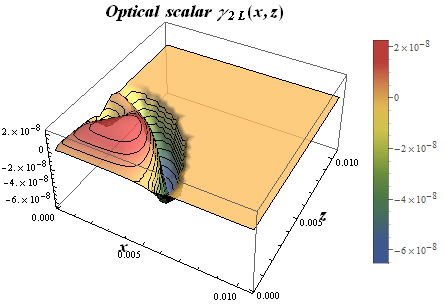}
	\caption{
It can be seen that both components of the shear show a rich
structure, that obviously disappear outside the void.
		Left: Real component of the shear $\gamma_{1L}$. 
		Right: Imaginary component of shear $\gamma_{2L}$.
	}
	\label{fig:shear1-shear2}
\end{figure}

One can see that the magnitude of the optical scalar functions for the case of typical
voids imbedded in a cosmological scenario
is very small in agreement  with previous analysis \cite{Amendola1999}.

It can be observed that the behaviour of the shear shows a rich structure, and that
neither the component nor the norm follow an expected quasi elliptical 
shape, coming from a projected spheroidal symmetric distribution.
Instead, when one looks at the behavior of the convergence $\kappa$,
 shown in figure (\ref{fig:kappa-shear}),
it can be seen a simpler behaviour.

\section{Final comments}

In this work on exact geometries  with spheroidal symmetry
which are suitable imbedded in a standard cosmological model,
we have presented a very broad class of metrics  useful for the 
description of over and underdensities matter distributions, namely 
equations (\ref{eq:ds-general-prolate-imbedded}) and (\ref{eq:ds-general-oblate-imbedded}).
These geometries are characterized by two `radial' functions 
which can be set
appropriately to give a R-W geometry in the range $ r \geqslant r_0$, for some prescribed fixed 
comoving radius $r_0$; while
the internal region $r < r_0$ is interpreted to represent an spheroidal system which grow
with the expansion factor $A(t)$.
The two scalars have a direct physical interpretation; the function $M(r)$ which was 
chosen such that in th limit $r_\mu \to 0$ coincides with the usual mass in spherical symmetry,
has also here a close relation with the energy density of the spheroidal distributions, 
as can be seen from inspection of figure \ref{fig:ProlateVoid}.
The function $F(r)$ has been used in the past to 
model peculiar spacelike components of the energy-momentum tensor; which are useful for 
the description of dark matter phenomena\cite{Gallo:2011hi}.

Using these exact geometries,
we have presented a model of void which admits within its interior 
region ($r < r_0$) an energy-momentum tensor which posses non-trivial components 
in general.
We have not found in the literature other presentations of exact metrics that
represent a cosmic void imbedded in an homogeneous cosmological 
Universe with spheroidal symmetry.
This model is at the same time an extension of a simple radial profile\cite{Amendola1999}
employed for spherical voids.
It has been shown that effectively it can be used as an appropriated model 
for representing underdensity distributions in a non-perturbative way.

As an interesting application we have taken the case of prolate spheroids for
performing a weak lensing analysis in the usual thin lens approximation.
Our calculation makes use of the techniques presented in \cite{Boero:2016nrd} for 
computing the optical scalars of the lens. 
In this example we have seen that although the expected signal is rather low,
some non-trivial features are displayed; 
for instance, the shear field does not recreate the expected quasi-elliptical 
geometry of a projected ellipsoid in the lens plane such as shown in section 
\ref{sec:weak}. 

We expect that the geometries presented here will be very useful in the 
study of systems requiring a more sophisticated modeling than the usual
spherical assumption; specially in the case of voids where the need 
for more general descriptions of their geometry seems very demanding.

In the future we plan to apply these type of models
to the description of specific observations of astrophysical systems.


\acknowledgments{We acknowledge support from CONICET, SeCyT-UNC and Foncyt.}

\providecommand{\href}[2]{#2}\begingroup\raggedright\endgroup

\end{document}